\documentclass[12pt]{aastex}
\usepackage{multirow}
\usepackage{graphics,graphicx}

\received{}
\begin{document}

\title{Chandra/ACIS Subpixel Event Repositioning. II. Further Refinements and Comparison between
Backside and Front-side Illuminated X-ray CCDs}

\author{Jingqiang Li\altaffilmark{1}, Joel H. Kastner\altaffilmark{1},
Gregory Y. Prigozhin\altaffilmark{2}, Norbert S.
Schulz\altaffilmark{2}, Eric D. Feigelson\altaffilmark{3},
Konstantin V. Getman\altaffilmark{3}}

\altaffiltext{1}{Chester F. Carlson Center for Imaging Science,
Rochester Institute of Technology, 54 Lomb Memorial Dr.,
Rochester, NY 14623; JL's email: jxl7626@cis.rit.edu}
\altaffiltext{2}{Center for Space Research, Massachusetts
Institute of Technology, Cambridge, MA 02139}
\altaffiltext{3}{Department of Astronomy and Astrophysics,
Pennsylvania State University, 525 Davey Laboratory, University
Park, PA 16802}

\begin{abstract}

We further investigate subpixel event repositioning (SER)
algorithms in application to Chandra X-ray Observatory (CXO) CCD
imaging. SER algorithms have been applied to backside illuminated
(BI) Advanced CCD Imaging Spectrometer (ACIS) devices, and
demonstrate spatial resolution improvements in Chandra/ACIS
observations. Here a new SER algorithm that is charge split
dependent is added to the SER family. We describe the application
of SER algorithms to frontside illuminated (FI) ACIS devices. The
results of SER for FI CCDs are compared with those obtained from
SER techniques applied to BI CCD event data. Both simulated data
and Chandra/ACIS observations of the Orion Nebular Cluster were
used to test and evaluate the achievement of the various SER
techniques.
\end{abstract}


\section{Introduction}
Subpixel event repositioning (SER) algorithms can be used to
improve the spatial resolution of Chandra X-ray imaging with the
Advanced CCD Imaging Spectrometer (ACIS), by reducing photon
impact position (PIP) uncertainties to subpixel accuracy.
Utilizing the extra information provided by the observation,
--- in particular, event charge split morphologies  and the
telescope pointing history --- SER techniques essentially change
the shape and decrease the size of the detector pixel. Therefore
the image quality degradation due to pixelization is reduced.
Tsunemi et al.\ (2001) first introduced SER methods for ACIS
imaging, describing a technique to reposition corner-split events.
In Li et al.\ 2003 (hereafter, paper I), SER algorithm
modifications were presented for back-illuminated (BI) devices, by
including single pixel events and 2-pixel split events to increase
the statistical accuracy as well as to improve detection
efficiency (from using $\sim$25\% of events to $\sim$95\% of
events). In this (static SER; hereafter SSER) formulation, the
repositioned event landing locations do not depend on energy.

Employing a high fidelity BI CCD model (Prigozhin et al.\ 2003),
we further modified SER by determining event PIPs according to
photon energies (energy-dependent SER; hereafter EDSER). Both CCD
simulations and real CXO observations demonstrate the improved
performance for SSER and EDSER compared to the Tsunemi et al. 2001
model (TSER), with EDSER displaying the best performance (Li et
al.\ 2003).

In paper 1 SER algorithms were discussed only for the BI devices.
Similar ideas can be applied to the FI devices, the details of the
implementation, though, are not same. The reason for that is that
photon absorption and charge spreading mechanisms differ
significantly for the two types of CCDs, especially at low X-ray
energies.

The collection of signal charge occurs near the front surface, the
same one that is illuminated by the incoming photons in the FI
CCD. Much larger fraction of photons interact close to the surface
of the device where electric potentials are influenced by the
grounded channel-stop layer, resulting in a very different charge
splitting pattern compared to the one in the BI devices. On
average charge clouds are formed closer to the collecting
potential wells and travel shorter distances, therefore having
less time to expand. Smaller charge clouds reduce the possibility
of forming split events.

A thicker dead layer covering vertical charge-splitting  pixel
boundaries of FI CCD is another factor contributing to reduction
of the share of split events. As a result TSER technique for FI
devices suffers seriously from low detection efficiency.

Mori et al. (2001) effectively modified TSER to SSER, by adding
single pixel and 2-pixel split events. However, they assume that
2-pixel events land on the center of split boundary, for both BI
and FI devices. In Paper I we showed that this assumption for
2-pixel split events impact position is inappropriate for BI CCDs,
and it follows that this assumption likely is not satisfactory
either for FI devices.

Here we describe modifications to SSER and EDSER algorithms for FI
devices. These modifications are based on a physical model of FI
CCDs (Prigozhin et al.\ 1998), as well as CXO observations with FI
ACIS CCDs. In addition, we describe a new SER technique that is
dependent on charge split proportion (CSDSER), and we apply this
method to both CCD types.

\section{Static SER for FI CCDs} \label{sec:sec2}

In the FI static SER method, as for static BI SER, single pixel
events and 2-pixel split events were added to corner split events,
in order to improve photon counting statistics. FI devices
generate far fewer corner split events, compared to BI devices
(see table \ref{branching}). Because the charge cloud has a
relatively small size, only photons that interact with silicon
close to boundaries result in split events in the case of FI
devices. Using detailed CCD model (Prigozhin et al.\ 1998b), we
simulated a distribution of events across the pixel and found that
for FI devices, single pixel events can occur almost everywhere
within a pixel except areas very close to corners and boundaries,
i.e., are constrained within an area only slightly smaller than a
CCD pixel. Two-pixel split events are generated by photons that
are absorbed in areas restricted to the pixel boundaries, while
the impact positions of corner split events are limited to the
diamond-shaped areas, diagonally oriented and heavily populated
towards pixel corners. Because the charge cloud size is very small
compared with ACIS pixel size, single-pixel events will have the
biggest position uncertainty in both dimensions, and corner split
events have the smallest uncertainty among all the events in both
dimensions. Two-pixel split events have relatively small landing
position uncertainties in the direction perpendicular to the split
boundary, and have uncertainties similar to those of single-pixel
events in the direction parallel to the split boundary. Thus,
properly repositioning both corner and 2-pixel split events will
essentially decrease the ACIS pixel size.

\begin{table}[!b]
\caption{Event branching ratios for CXO Orion Nebula Cluster
  observations$^{a}$.}\label{branching}
 \begin{center}
  \begin{tabular}{|c|c|c|c|c|}
  \hline
  ACIS CCD type & Corner-split events & 2-pixel split events & Single pixel events& Total split events \\
  \hline
  BI$^{b}$ & 20.8-38.8\% & 38.4-50.3\% & 10.9-25.2\% & 69.6-83.8\%\\
  \hline
  FI$^{c}$ & 1.0-5.7\% & 15.0-23.3\% & 70.5-83.3\% & 16.5-29.0\%\\
  \hline
  \end{tabular}
  \end{center}

  Notes.---\\
  a). Table values are calculated from 20 and 32 individual bright sources with Gaussian
  shapes for BI and FI observations, respectively.\\
  b). CXO ObsID 04.\\
  c). Chandra Orion Ultrdeep Project data (see sec. \ref{sec:sec4})

\end{table}
In our initial FI SER implementation, we assume that corner split
events take place at the split corners instead of event pixel
centers, and 2-pixel split events occur at the centers of split
boundaries, 0.47 pixel away from the pixel centers. Single pixel
event PIPs remain at the event pixel centers. Note that the 0.47
pixel offset for 2-pixel split events here is different from BI
SSER, in which the shift is 0.366 pixel. These 2-pixel split event
shifts for FI and BI SSER were determined from FI and BI CCD model
simulations, respectively. As in Paper I, we refer to this
modified algorithm as {}``static'' (energy-independent) SER. The
algorithm's schematics can be found in Figure 1 of Paper I.

\begin{figure}
{\centering \resizebox*{3.5in}{!}{\includegraphics{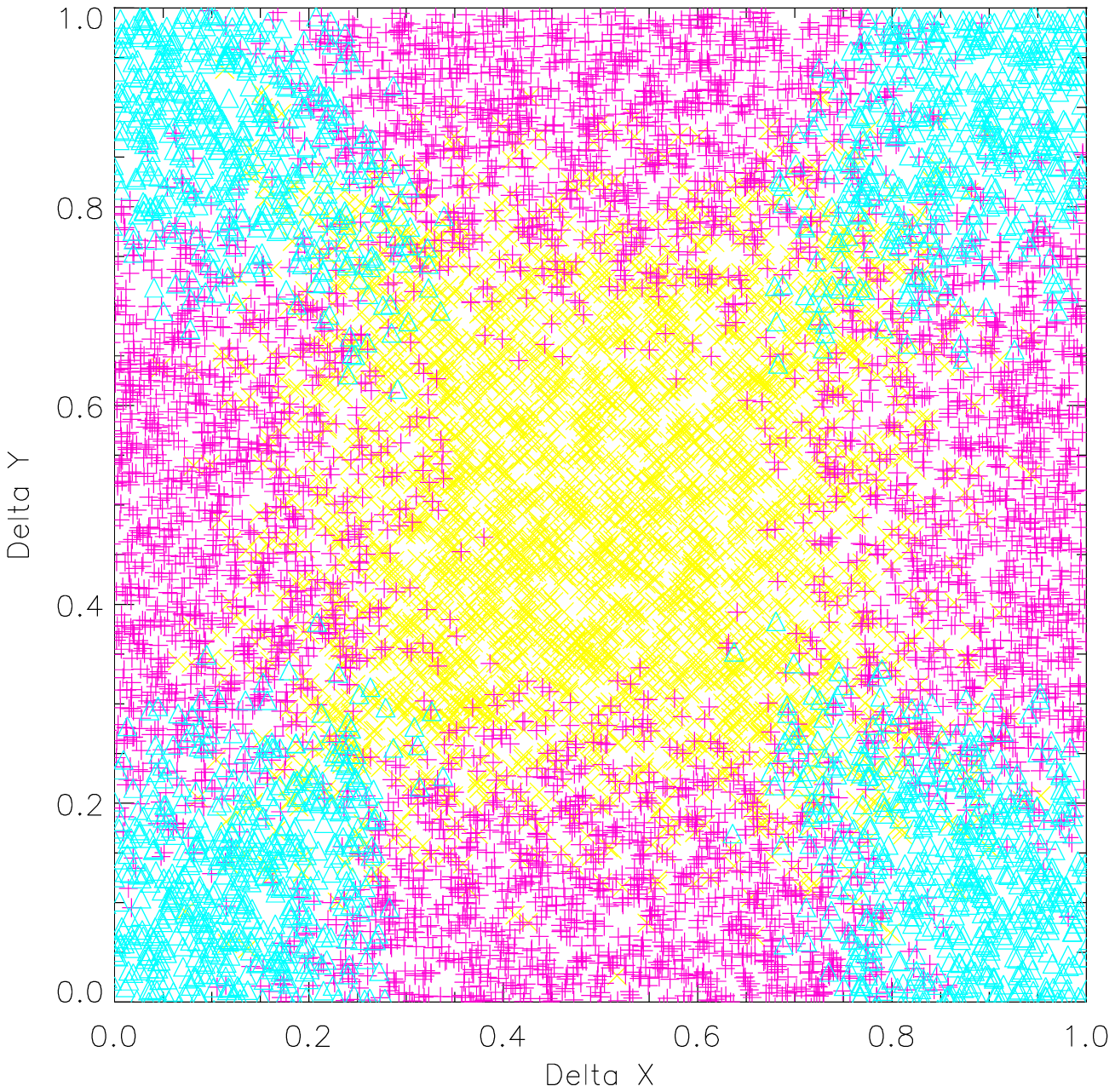}}
\par \par}
{\centering \resizebox*{3.5in}{!}{\includegraphics{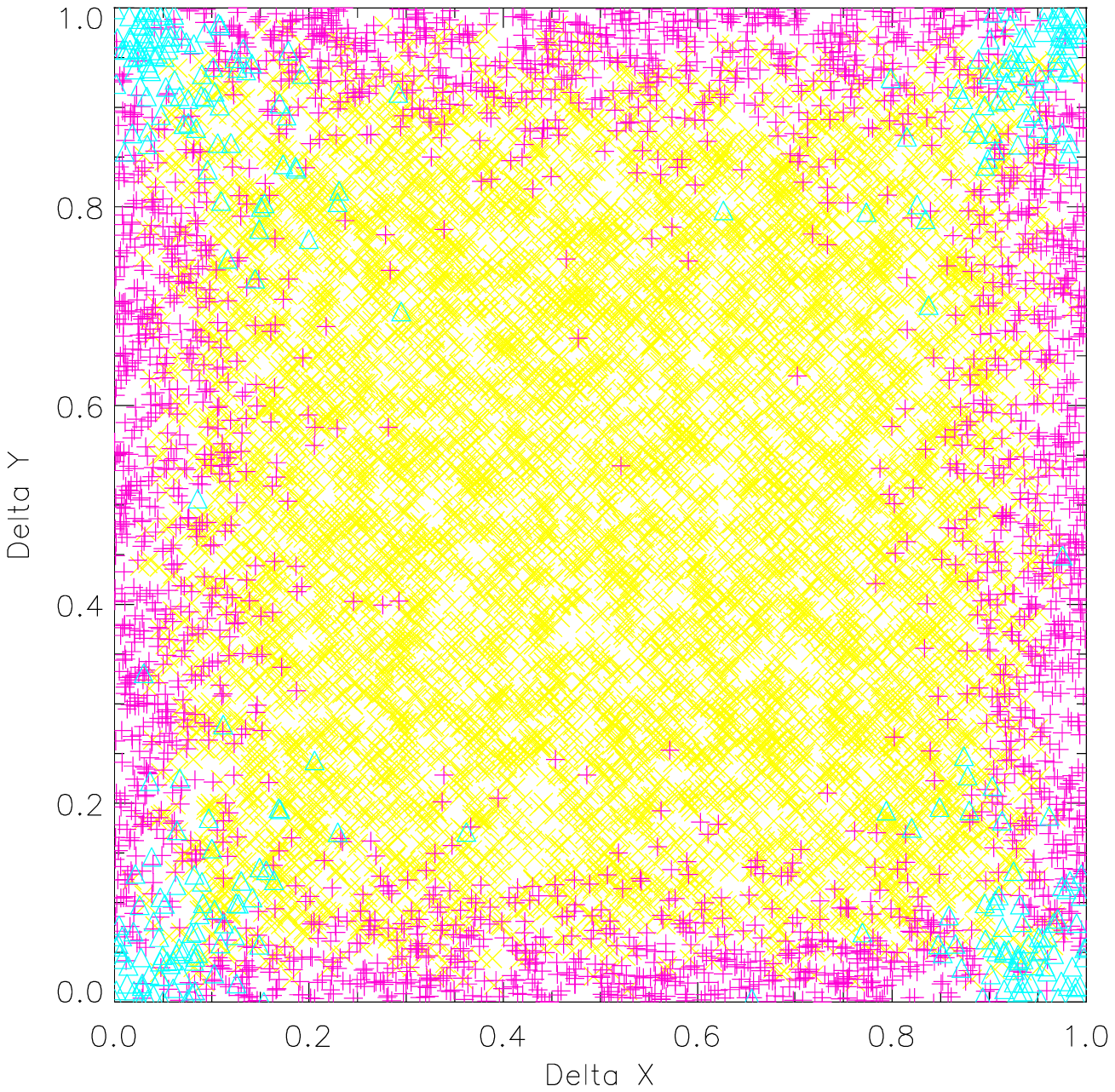}} \par} 
\caption{The photon impact positions for 3
subgroups of 13 {}``viable'' event grades for BI (top) and FI
(bottom) devices. Plus signs stand for the PIPs of 2-pixel events
within a pixel, while triangles represent the PIPs of corner (3-
or 4- pixel) split events. The crosses are the PIPs of single
pixel events. All the photons have energy of 1.74 keV.}
\label{splitevent}
\end{figure}

Simulations for BI CCDs show that a given type of event can be
formed in a fairly large area, and the mean offset of this event
type from pixel center is determined by the charge cloud size,
which is energy dependent. The same principles hold for FI devices
too, but differences exist; in particular, split events are much
less probable and occur closer to split boundaries, as shown in
Figure \ref{splitevent} for 1740 eV
photons. 

\begin{figure}
{\centering \resizebox*{7in}{!}{\includegraphics{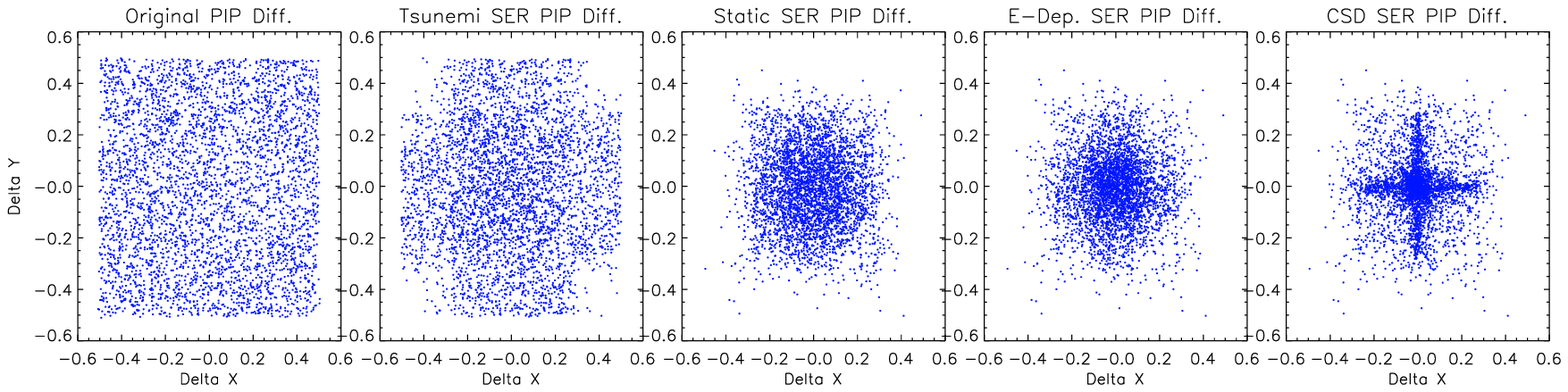}}\par \par} {\centering
\resizebox*{7in}{!}{\includegraphics{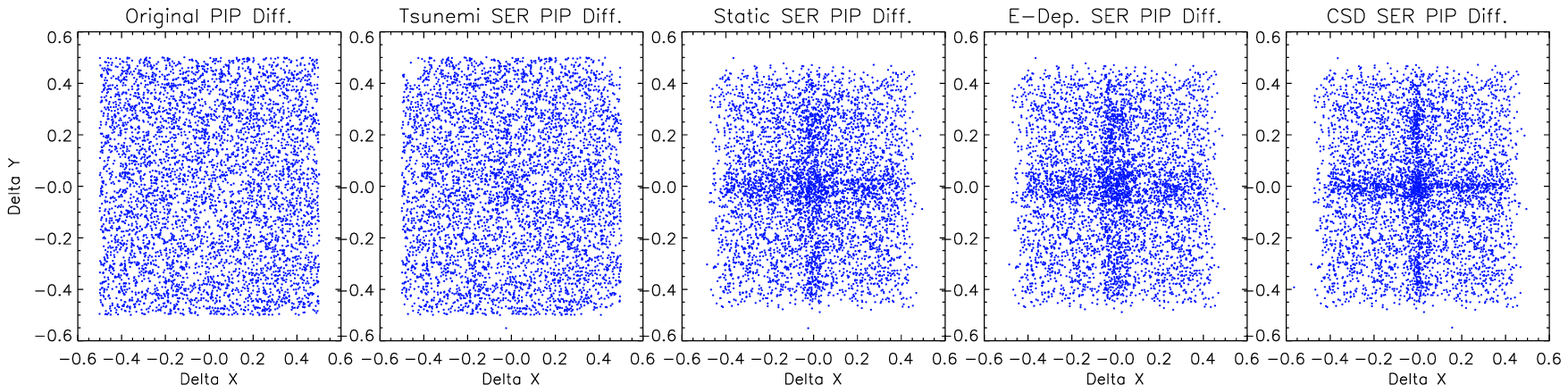}}\par}

\caption{Differences between actual photon impact positions and
processed event assumed locations for 1.74 keV events, in Chip
coordinates. $1^{st}$ panel (from left): ACIS assumed PIP;
$2^{nd}$ panel: correction using corner events only (Tsunemi et
al. 2001); $3^{rd}$ panel: static SER correction; $4^{th}$ panel:
EDSER correction; $5^{th}$ panel: CSDSER correction. The panels
are in units of pixels. The top row panels are for BI devices,
while the bottom row panels are for FI devices, for 4000 (BI) and
5000 (FI) photons with uniformly random landing positions.}
\label{half_shift_compare}
\end{figure}

The first three panels in Figure \ref{half_shift_compare} show the
improvement in determination of PIPs enabled by the modified SER,
using FI CCD simulated data at an energy of 1740 eV. For
comparison, the BI simulated data of same energy photons (shown in
Paper I) is included. In each panel, we show the differences
between actual PIPs and repositioned PIPs from various models in
chip coordinates, for all three subgroups \footnote{Note the
difference between three subgroups of events and three subgroups
of \emph{split} events. The three subgroups of events mean single
pixel events, 2-pixel split events, and corner (including 3- and
4-) split events. The three subgroups of split events represent
2-, 3- and 4-pixel split events.} of events. The plot axes are in
ACIS pixel units, i.e., 0.5 difference represents 12 $\mu$m, and
indicates photons that interacted near the pixel boundaries. The
first panel from left shows the difference of actual PIPs for a
random spatial distribution of events with unrandomized,
standard-processed PIPs which are assumed to lie at the event
pixel centers; one can see the expected uniform random
distribution within the pixel. The second panel is the difference
after applying TSER, in which only corner split events were
repositioned. A big improvement for the small fraction of events
that occur near corners can be seen. However, due to the small
proportion of corner split events, there is no correction for most
events. This fact is more obvious for the FI simulations. The
third panel shows the difference after the static SER correction,
in which the 2-pixel split events also were repositioned. For FI
devices, SSER results in a ``\#''-shape structure, because the
uncertainty of 2-pixel events can only be minimized in one
direction. However, the smaller PIP differences of the SSER method
relative to the Tsunemi et al.\ (2001) method are apparent, with
the improvement more obvious for BI devices. The other two panels
in the Figure will be discussed later.

Essentially, the PIP differences plotted in figure
\ref{half_shift_compare} represent the corrected PIP uncertainty
(or probability distribution) within a pixel, and therefore can be
considered as representing the ACIS pixel shape and size after SER
correction. Adopting this concept, one sees that the far left
panel reflects the ACIS pixel after standard CXO/ACIS processing,
i.e., a square pixel with 24 $\mu$m width. After TSER and SSER
correction, the effective ACIS pixel becomes smaller in size, and
no longer has uniform spatial response.

\clearpage

\section{Further Modifications to SER Based on ACIS CCD Simulations}
\label{sec:sec3}
\subsection{Energy-dependent SER for FI CCDs}
\begin{figure}
{\centering \resizebox*{4in}{4in}{\includegraphics{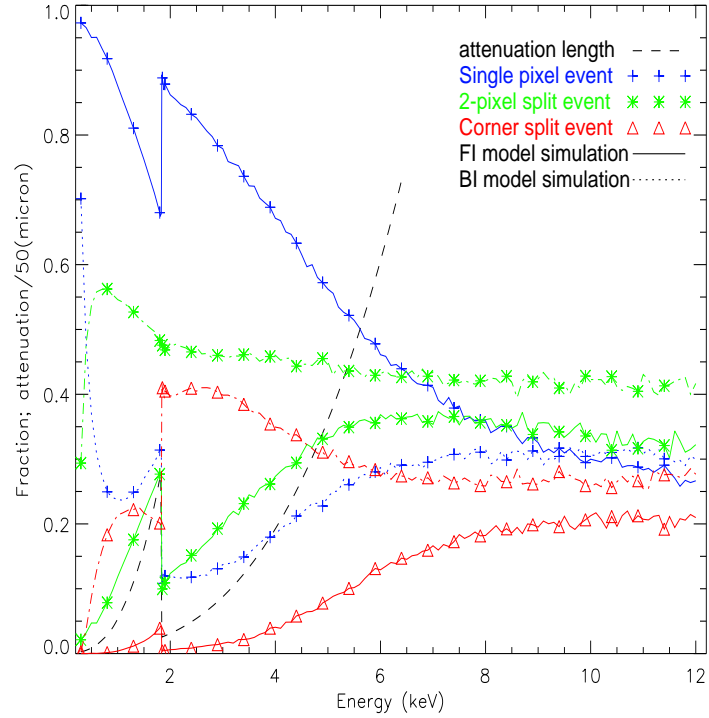}}
\par}
\caption{The fraction of different event grades versus photon
energy. Results from simulations of FI (solid line) and BI (dotted
line) CCD model. The X-ray attenuation length in silicon is
overplotted, in units of 50 microns. } \label{split_percentage}
\end{figure}

\begin{figure}
{\centering \resizebox*{4in}{4in}{\includegraphics{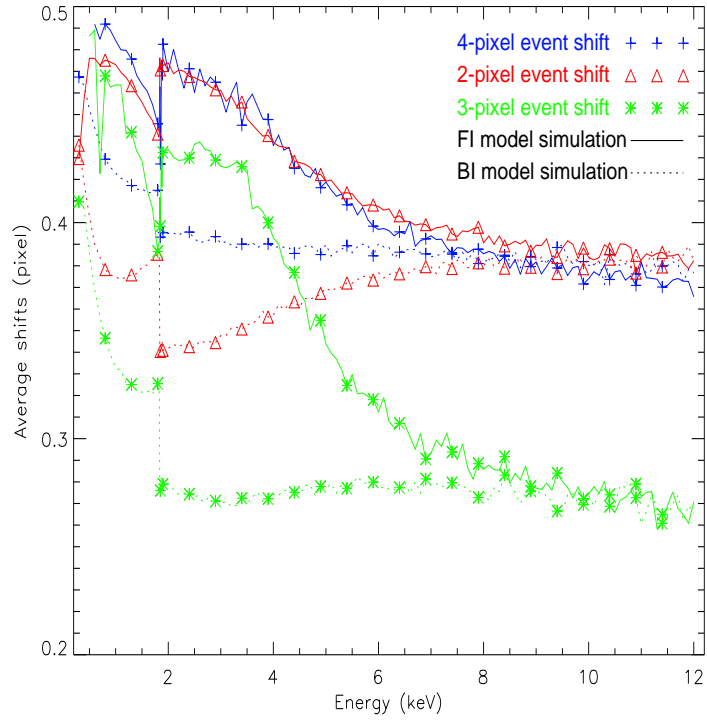}} \par}

\caption{The mean shifts from pixel centers of the 3 subgroups of
split events, according to the photon energy. FI CCD model
simulations are plotted with solid line, while BI simulations with
dotted line.} \label{shift_energy}
\end{figure}

In Paper I, an energy-dependent SER method was proposed for BI
devices, based on simulations for a BI CCD model. The advantages
of this method were demonstrated from both simulated data and real
observations. Figures \ref{split_percentage} and
\ref{shift_energy} show the motivation for ED SER from the
simulation, i.e., the branching ratio and the mean offset of the
split event position, and, hence, the mean shifts for each
subgroup of split events are strongly energy dependent. Therefore,
adjusting assumed PIPs according to energy should significantly
improve SER performance.

Figure \ref{split_percentage} shows the percentage of events of a
given split morphology as a function of photon energy, while
Figure \ref{shift_energy} shows the mean shift in position for
different split event types, for FI devices. For comparison, we
include the same plots for BI CCDs that were published in Paper I.
Note the differences between BI and FI devices. For BI CCDs, both
subgroup event percentage and mean PIP shift depends sensitively
on energy, at low energy (E $<$ 2 keV). The 3 subgroups of split
event percentages and PIP shifts are insensitive to energy for E
$>$ 6 keV. This reflects the fact that, for photons with energy
exceeding 6 keV, the characteristic penetration depth becomes
comparable to or larger than the thickness of the ACIS BI CCD,
which is only 45 microns. In contrast, ACIS FI CCDs are much
thicker, with larger depletion depth ($\sim$ 70 $\mu$m, Prigozhin
et al.\ 1998a). Therefore, the branching ratios and PIP shifts
depend sensitively on energy over most of the CXO/ACIS bandwidth.

EDSER consists of repositioning the split event PIPs by event
grade, using the mean PIP offset look-up table as a function of
photon energy derived from data shown in figure
\ref{shift_energy}. PIP determination benefits from applying the
mean energy-dependent shifts for different split event groups. The
fourth panels (from left) of Figure \ref{half_shift_compare}
demonstrate the PIP differences after EDSER for BI and FI devices.
Compared with static SER, the EDSER BI data displays a more
concentrated structure in the center, indicating the split events
were relocated more accurately, and the energy dependent SER
method will improve SER performance, via better PIP determination.
However, due to narrower confinement of split events to pixel
boundaries, one doesn't see the same improvement for FI data in
Figure \ref{half_shift_compare}, indicating that EDSER may not
yield much gain over SSER, for FI CCDs.

\clearpage
\subsection{Charge Split Dependent SER}
\begin{figure}
{\centering \resizebox*{6in}{!}{\includegraphics{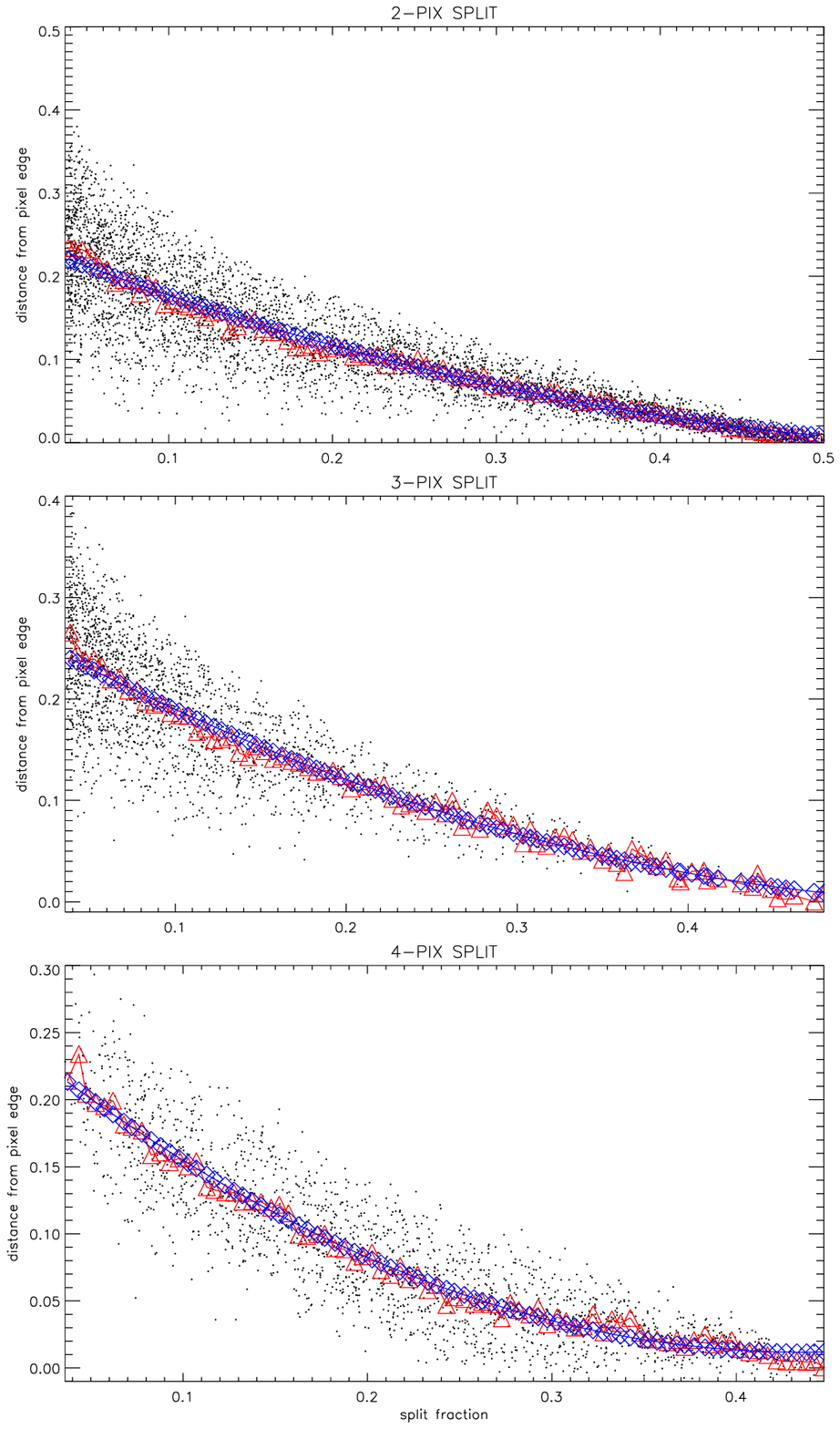}
{\includegraphics{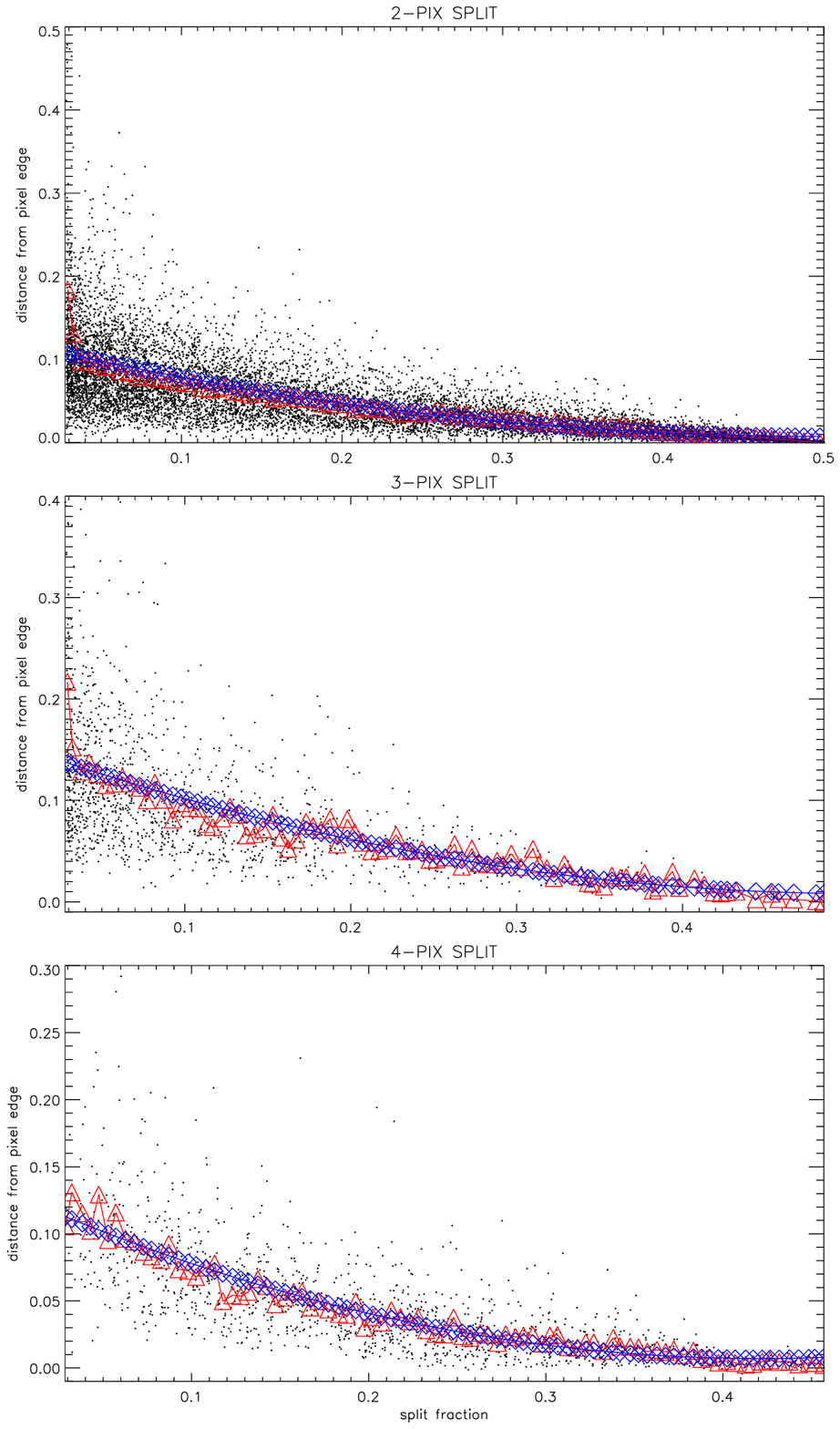}} \par}}

\caption{The distance of photon landing locations from the split
  boundaries as a function of charge split proportion for
  three split event types. The photons have energies of 1740 eV.
  Simulations were performed with MIT BI (left column) and FI (right column) ACIS models.
  The dots in the panels represent the PIP relative to split
  boundaries, while the red triangles are the local averages of the PIPs, and the
  blue lines are the polynomial regression curves of the local
  averages.} \label{ECSPD}
\end{figure}

Simulations show that, for a split event, the proximity to the
split boundary of a photon impact position is related to the
proportion of the charge deposited in split pixels relative to the
total charge generated by the photon. This fact provides
motivation for an SER algorithm that is both energy and charge
split proportion dependent. Figure \ref{ECSPD} shows distances of
PIPs (relative to split boundaries) as a function of the charge
split proportion, for three types of split events. ACIS CCD models
were used for these simulations, at a photon energy of 1740 eV.
The simulated results are shown in the left and right columns, for
BI and FI devices respectively. The measured fraction is the
proportion of charge within a split pixel relative to the total
charge generated by the event, including all split charge that
exceeds the split threshold. For 3- and 4-pixel split events, the
charge fraction in both horizontal and vertical split pixels was
measured independently. The charge fraction in the diagonal split
pixel of the 4-pixel split events was not measured, since the
fractions from the other two split pixels already provide
information about photon landing locations.

The plot shows that, with only energy information, the PIP
uncertainty is relatively big since it includes all ``local''
uncertainties. By including charge split proportion information,
one can divide the uncertainty into local uncertainties, i.e., the
uncertainty at each split fraction. For example, for a 3-pixel
split in BI device (the middle panel of the left column), the
total uncertainty is about 0.4 pixel, while the local uncertainty
at 0.4 split fraction is only about 0.03 pixel. Therefore,
including charge split information, CSDSER will greatly reduce PIP
uncertainties. The function describing PIP offset in terms of
split charge fraction for horizontal and vertical directions is
assumed indistinguishable\footnote{Even though the pixel physical
boundaries are different in the two perpendicular directions,
i.e., one boundary is provided by channel stops, while the other
is caused by the gate with lower voltage, CCD simulations don't
show obvious split property differences for these different
boundaries.}, for a given split-event subgroups, at the same
energy.

The rightmost panels in figure \ref{half_shift_compare} show the
simulated PIP uncertainties after CSDSER correction for BI and FI
devices, at the energy of 1740 eV. The improvement in PIP
determination for those panels can be seen, especially for BI
devices, compared with EDSER correction. Figure
\ref{half_shift_compare} suggests an increasing degree of image
quality improvement can be achieved, by using SSER, EDSER and
CSDSER.

\clearpage


\clearpage
\section{Testing SER: End-to-end Simulations}\label{sec:sim}
MARX (Model of AXAF Response to X-rays) is a program suite that
can run in sequence to simulate Chandra on-orbit performance, with
FITS file and image output (MARX technical manual). The built-in
instrument models, including HRMA (High Resolution Mirror
Assembly) and focal plane detectors, enable MARX to perform a
ray-trace and thereby simulate Chandra CCD imaging spectroscopy of
a variety of astrophysical sources. Post-processing routines can
simulate aspect movement and ACIS photon pile up. However, ACIS
simulations within recent versions of MARX do not include a
high-fidelity CCD model. Therefore, SER related simulations must
rely on CCD models (such as those described in sections
\ref{sec:sec2} and \ref{sec:sec3}) to analyze the CCD charge
distribution and event grade formation and, therefore, SER
implementation.

\begin{figure}
{\centering\resizebox*{4in}{!}
{\includegraphics{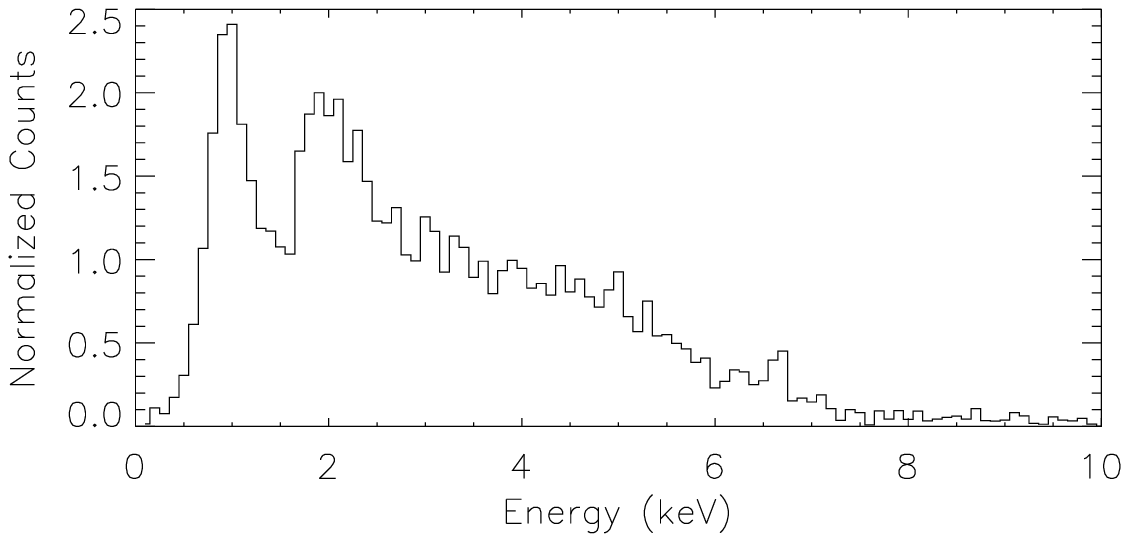}} \par}
{\centering
\resizebox*{4in}{!}{\includegraphics{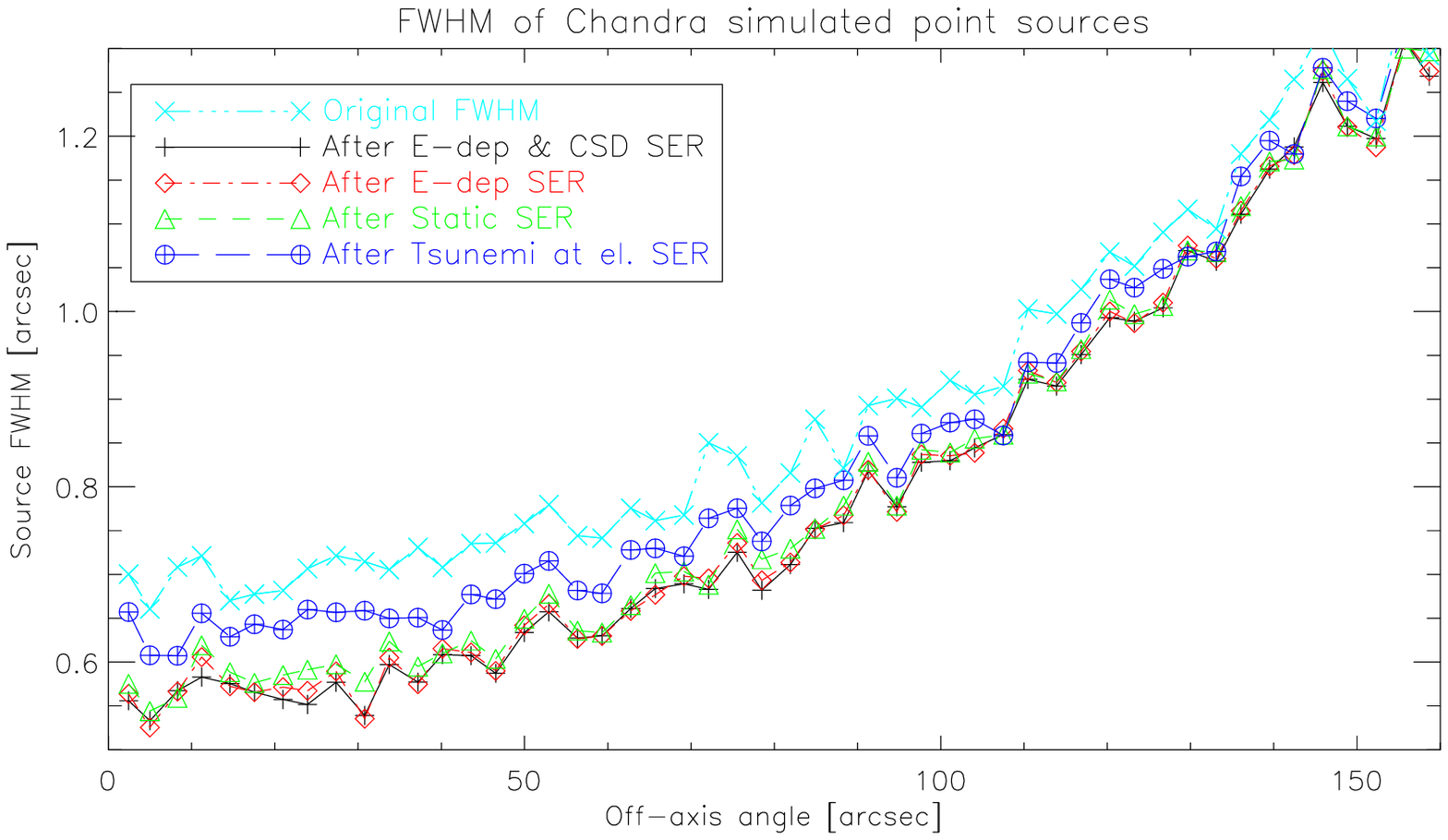}} \par}
{\centering
\resizebox*{4in}{!}{\includegraphics{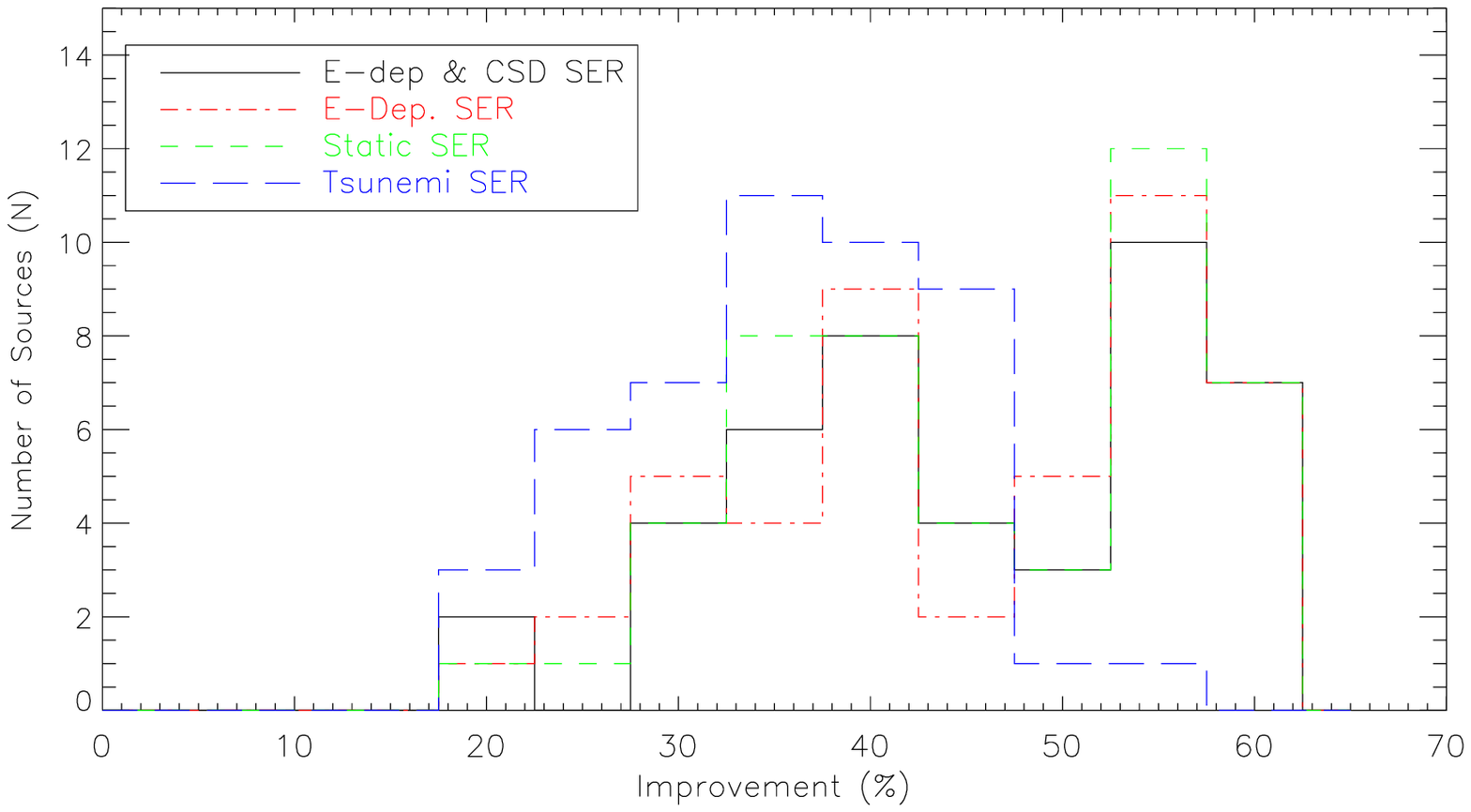}} \par}
\caption{Improvement comparison of different BI SER algorithms,
from BI CXO/ACIS point-source simulations. Top: spectrum used in
the simulations (see text). Middle: improvement in source FWHM as
function of source off-axis angle. Bottom: histogram of FWHM
improvements.} \label{simulation}
\end{figure}

\begin{figure}
{\centering
\resizebox*{4in}{!}{\includegraphics{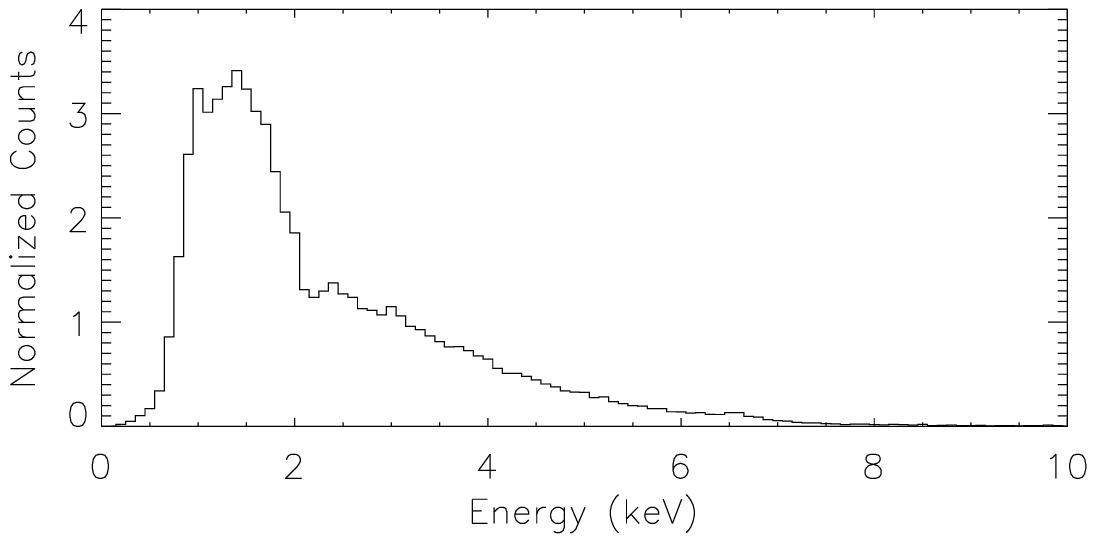}} \par}
{\centering
\resizebox*{4in}{!}{\includegraphics{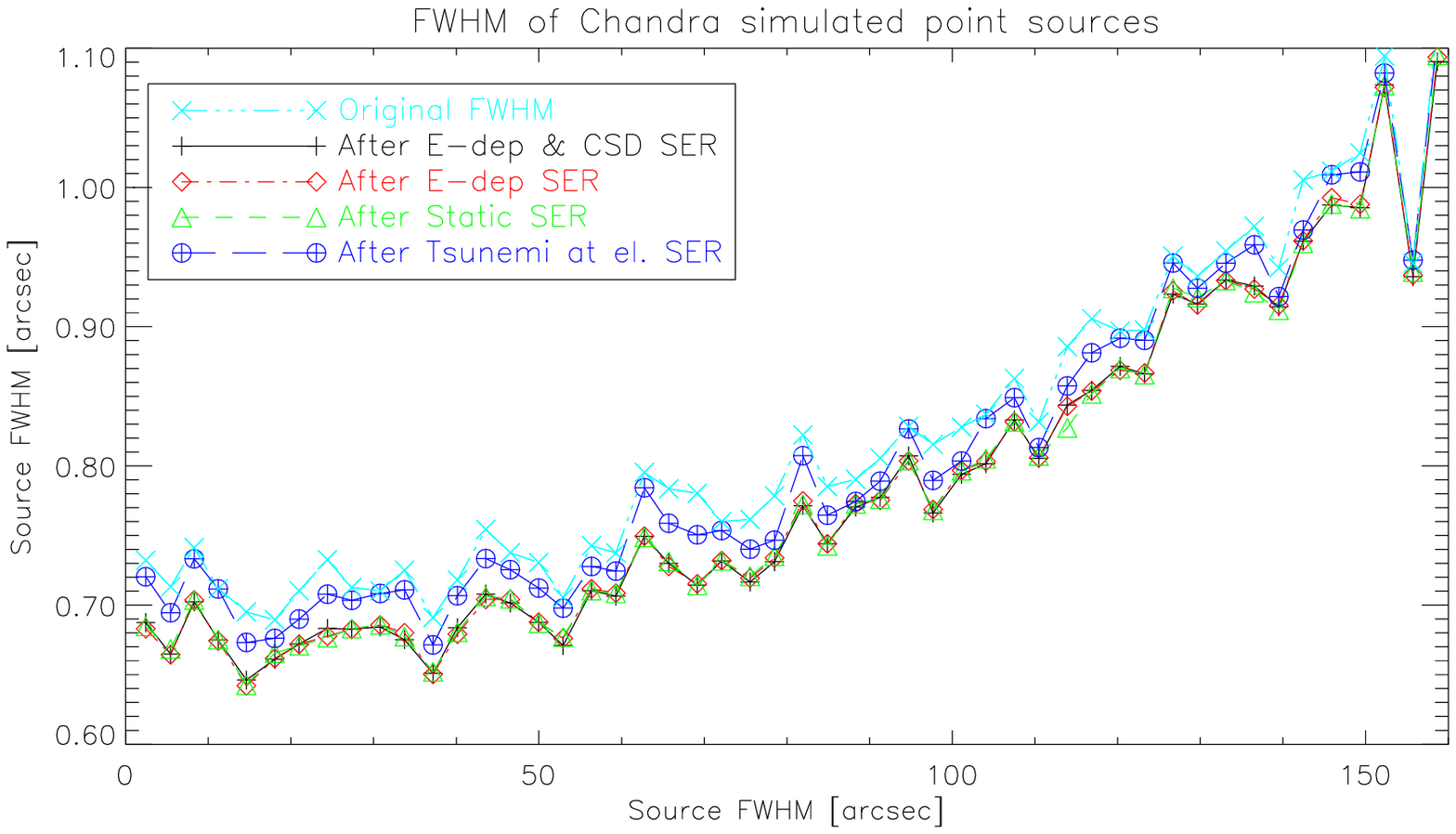}} \par}
{\centering
\resizebox*{4in}{!}{\includegraphics{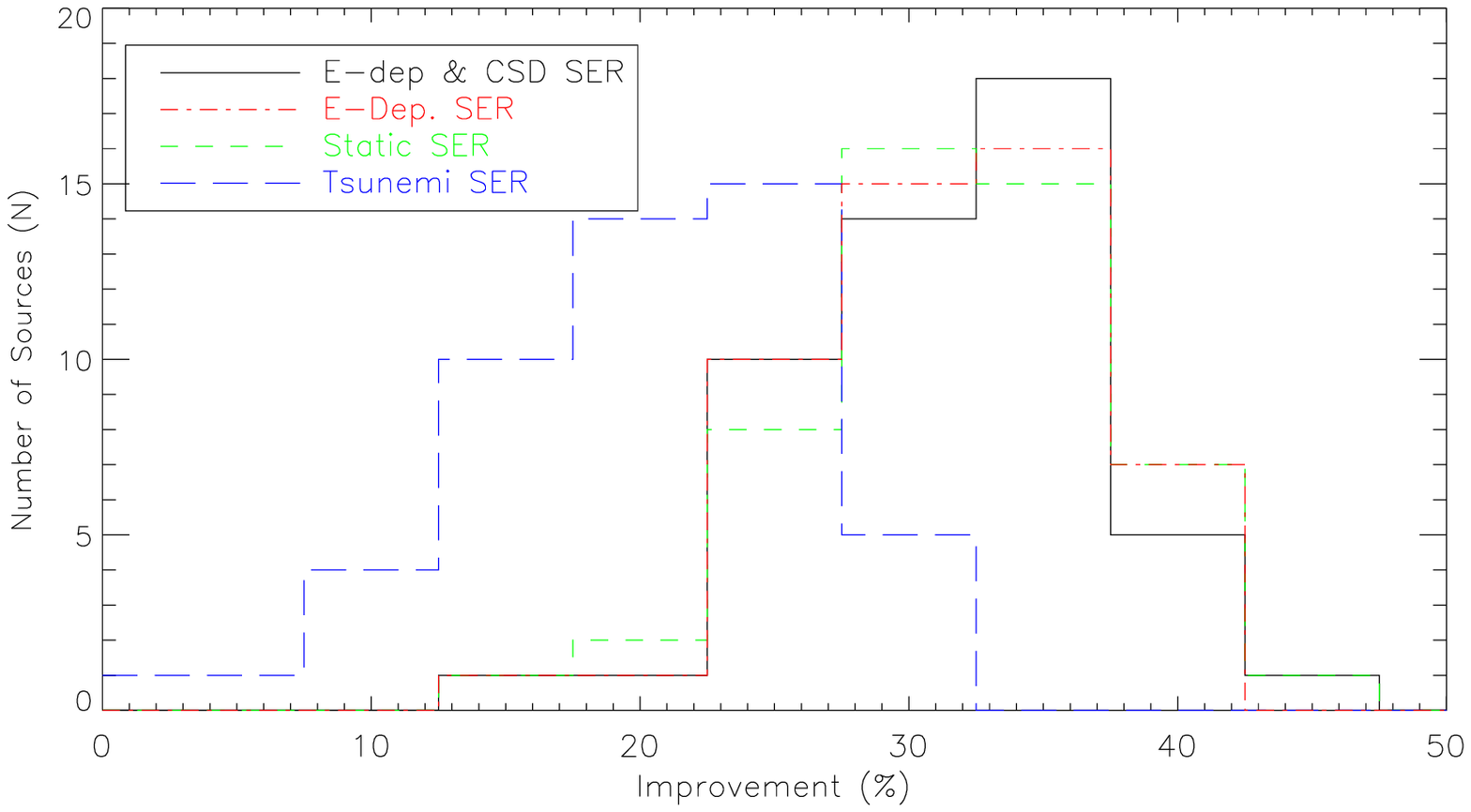}} \par}
\caption{As in Fig. \ref{simulation}, improvement comparison of
different FI SER algorithms from FI CXO/ACIS point-source
simulations.} \label{simulationFI}
\end{figure}

To test the extent to which various SER techniques should improve
image performance for BI and FI CCDs, we have carried out
simulations combining MARX with MIT BI/FI CCD models. We performed
simulations of 50 point sources with realistic spectral
distributions at positions ranging from on-axis to $160''$
off-axis, in steps of $3''.2$ (see figure \ref{simulation} and
\ref{simulationFI}). The MARX telescope ``internal dither'' model
was used, with standard (default) values of 1000 and 707 second
dither periods in RA and DEC directions, respectively, and an 8
arcsec dither amplitude in both directions.

The BI and FI simulation spectra are based on the averaged spectra
of BI and FI ONC observations, respectively. The BI simulation
spectrum was calculated from 20 point-like sources\footnote{The
sources were listed in table 1 of Paper I, excluding source 5 and
6, which may be affected by pileup.} of BI ONC (obsID 4), while
the FI simulation spectrum was calculated from an average spectrum
of 32 well-shaped X-ray sources from a deep CXO/ACIS-I observation
(see section \ref{sec:sec4}). These 32 sources were also used in
sections \ref{sec:sec2} and \ref{sec:sec4}, for source split event
branching ratios (table \ref{branching}) and various SER method
evaluations (figures \ref{fwhm_imp} and \ref{improve}),
respectively. The BI and FI spectra are plotted in figure
\ref{simulation} and \ref{simulationFI}, respectively.

The degree of improvement due to the SER algorithms was evaluated
numerically by calculating source FWHM before and after applying
SER. Tsunemi et al.\ (2001) and Paper I give the definition of
improvement; i.e., by assuming that {$\mit F_{B}$} and {$\mit
F_{A}$} are the FWHMs of a source before and after applying SER,
respectively, then the improvement $\Delta$ is defined as:
\begin{displaymath}
\Delta\,=\,\sqrt{{\mit F_{B}^{2}\,-\,F^{2}_{A}}}/F_{B}
\end{displaymath}

The results of the simulations are shown in Figures
\ref{simulation} and \ref{simulationFI} for BI and FI models,
respectively. 
The progressively better performance of SSER , EDSER, and CSDSER
is apparent in BI simulations, as expected, due to better PIP
determination from addition photon and charge split information
(see table \ref{res_comp}). However, the performance of SSER,
EDSER and CSDSER is very comparable in the case of FI devices,
even though we might expect to see the improvement (e.g., of
CSDSER relative to SSER) theoretically. In comparison to BI
devices, the lack of improvement in imaging performance under the
refined SER approaches for FI CCDs is most likely due to the
following factors:
\begin{enumerate}
    \item FI devices generate fewer split events than BI
    devices, especially corner split events.
    Therefore single pixel events dominate over the better
    repositioned split events.
    \item For soft sources, such as those simulated here, the charge cloud size is relatively
    small. Therefore most split events in FI CCDs are very close to the split
    boundaries, not widespread as in BI devices. As a result, the positional
    uncertainties of two-pixel split events forms a long arm cross
    structure after applying SER. The uncertainty in the direction parallel to the split
    boundary is larger and remains unchanged.
    \item Because of the small charge cloud, the PIP determinations of
    EDSER and CSDSER do not provide significant advantages  over the static
    method, as for BI devices.
    \item The slight potential improvement offered by CSDSER is degraded by telescope PSF,
    which includes contaminations from both HRMA PSF and aspect blurring.
\end{enumerate}
\begin{table}
  \begin{center}
  \caption{Comparison of various BI and FI SER
  methods.}\label{res_comp}
  \begin{tabular}{|c|c|c|c|c|c|}
   \cline{4-6}
   \multicolumn{3}{c}{} & \multicolumn{3}{|c|}{Degree of improvement}\\
   \hline
   & Off-axis range & CCD type & CSD $>$ ED$^{a}$ & CSD $>$ SSER$^{b}$ & ED $>$ SSER$^{c}$\\

   \hline
   \multirow{2}{20mm}{50 sources} & \multirow{2}{20mm}{0---$158''.7$ } & BI & $68\%$ & $86\%$ & $80\%$ \\
   \cline{3-6}
   & & FI  & $56\%$ & $52\%$ & $56\%$\\
   \hline
   \multirow{2}{20mm}{25 sources} & \multirow{2}{20mm}{0---$78''.5$} & BI & $64\%$ & $96\%$ & $88\%$ \\
   \cline{3-6}
   & & FI & $56\%$ & $60\%$ & $64\%$ \\
   \hline
    \end{tabular}
    \end{center}

  Notes. ---\\
  a). Percentage of sources for which CSDSER FWHM improvement is
  larger than that of EDSER.\\
  b). Percentage of sources for which CSDSER FWHM improvement is
  larger than that of SSER.\\
  c). Percentage of sources for which EDSER FWHM improvement is
  larger than that of SSER..

\end{table}

Figures \ref{simulation} and \ref{simulationFI} show that SER
algorithms are highly source location dependent, i.e., all SERs
have better performance for on-axis sources, and the improvement
decreases when off-axis angle increases. This is because the
telescope PSF increases in size with off-axis angle and,
therefore, the influence of the event repositioning decreases.


\clearpage
\section{Application of the SER Algorithms to X-ray sources in Orion}
\label{sec:sec4}

The steps involved in implementing SER on real Chandra
observations were discussed in paper I, and BI SER algorithms
(TSER, SSER and EDSER) were evaluated from data obtained for the
ONC (Orion Nebula Cluster, Schulz et al.\ 2001). Here we compare
CSDSER with other SER algorithms for BI data (top panels in
figures \ref{fwhm_imp} and \ref{improve}). The same implementation
steps hold for FI CCDs, except the chip orientation differences
within eight FI chips. Similar plots for applications of SER
methods on FI Chandra Orion Ultradeep Project (COUP) data are
shown in the bottom panels.

The Chandra Orion Ultradeep Project combines six consecutive
observations of the Orion Nebula Cluster taken in January 2003
with the Advanced CCD Imaging Spectrometer on board the Chandra
X-ray Observatory (Weisskopf et al.\ 2002). The total exposure
time of $0.84$ Ms and over 1600 sources are detected.

COUP data reduction started with the Level 1 event files provided
by the Chandra X-ray Center. Only events on the four CCDs of the
ACIS I-array were considered.  Event energies and grades were
corrected for charge transfer inefficiency (CTI) using the
procedures developed by Townsley et al.\ (2001 ApJL).  The data
were cleaned from a various potential problem events with the
grade, status, and good-time intervals filters as described in the
Appendix of Townsley et al.\ (2003).  Sequences of single pixel
cosmic ray afterglow events were identified but not removed from
the dataset at this time.  Bad pixel columns with the energies $<
700$ eV and the background events with the energies $> 10500$ eV
were removed.

Event positions were adjusted slightly in three ways.  First,
individual corrections to the absolute astrometry of each of the
six COUP exposures was applied based on several hundred matches
between a preliminary catalog of Chandra sources and near-infrared
sources in a forthcoming catalog from the ESO Very Large
Telescope.  Second, the sub-arcsecond broadening of the PSF
produced by the Chandra X-ray Center's pipeline randomization of
positions was removed. Third, the tangent planes of five COUP
exposures were re-projected to match the tangent plane of the
first observation (ObsID 4395).  The six exposures were then
merged into the single data event file used in this paper.

\subsection{Results}

\begin{figure}
{\centering \resizebox*{4in}{!}{\includegraphics{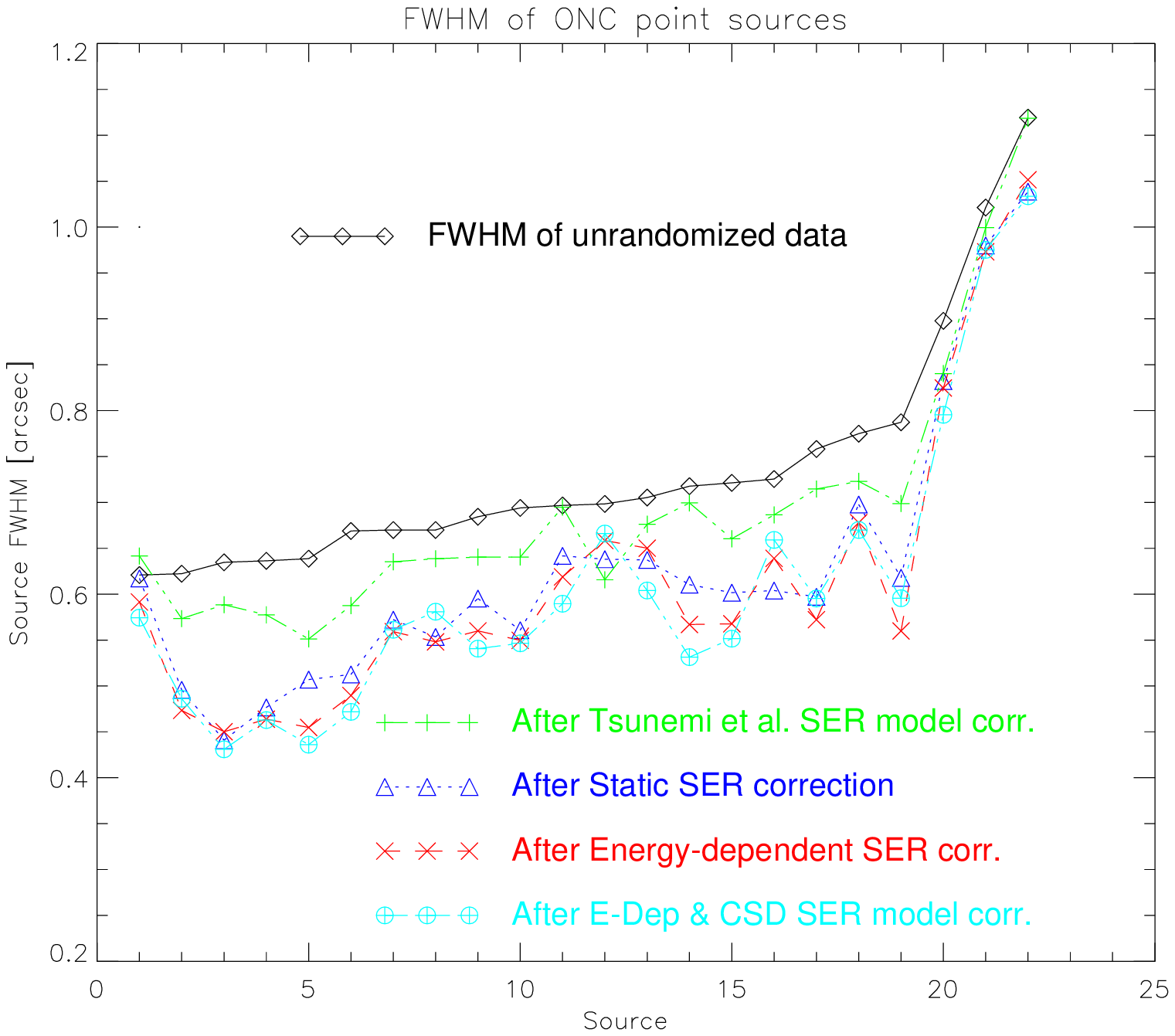}}
\par}
{\centering \resizebox*{4in}{!}{\includegraphics{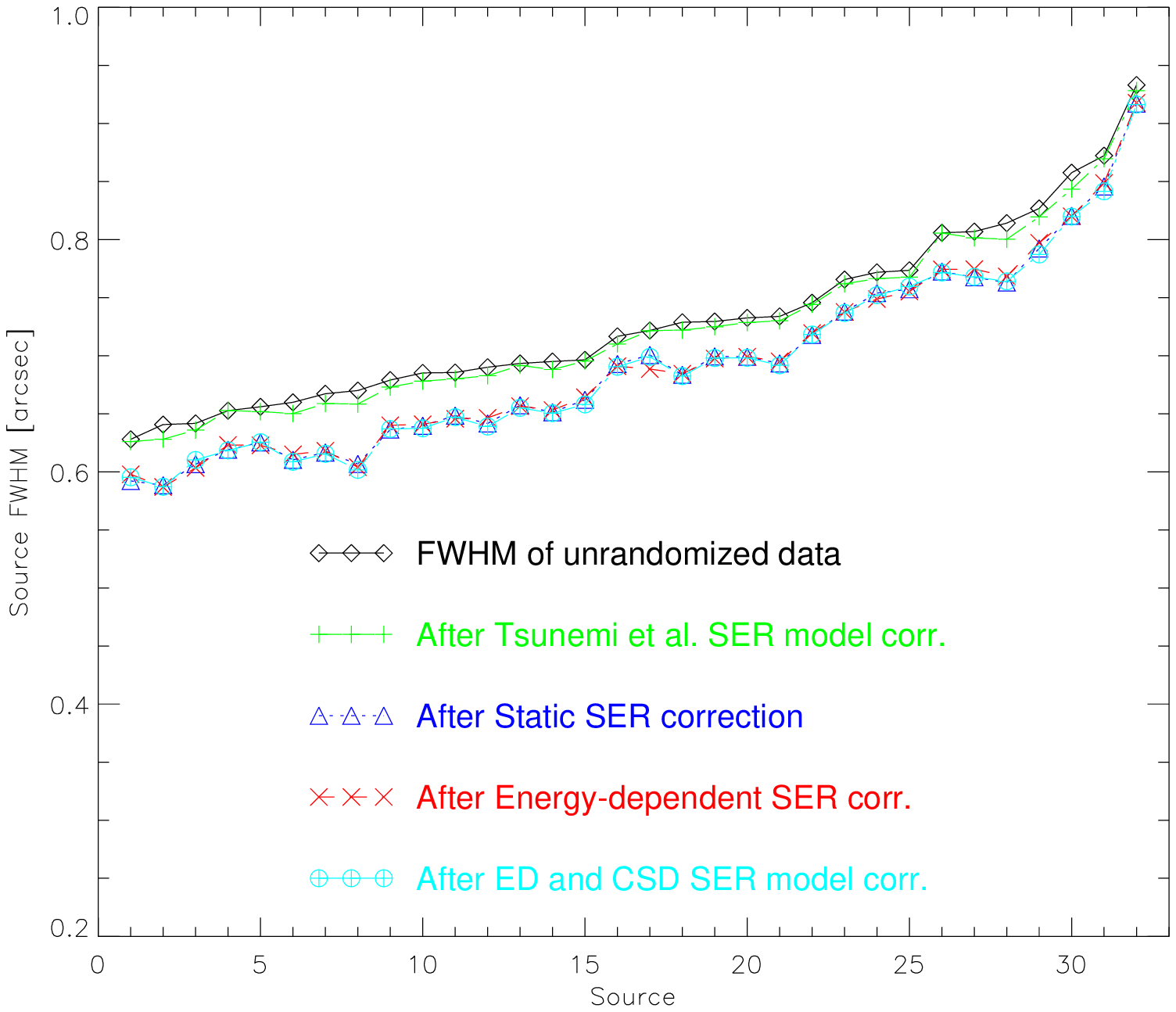}}
\par}
\caption{FWHM of BI and FI ONC point like sources before and after
applying various SER algorithms described in this paper.}
\label{fwhm_imp}
\end{figure}

Based on the above steps, we have plotted the FWHM of 22 bright
point-like sources in BI ONC data, obtained by Chandra/ACIS-S3.
The sources were selected to represent a range in off-axis angle
from $2''.72$ to $136''.8$, and in count rate from $0.0052$ to
$0.2791$ s$^{-1}$ (Paper I). The top panel in figure
\ref{fwhm_imp} shows that after applying SER technique to these
data, all SER algorithms (sec. \ref{sec:sec3}) improved the FWHM
for every source (except that source 1 has no improvement after
applying the Tsunemi et al.\ [2001] method). The bottom panel
displays 32 point-like sources (could be different with BI
sources) chosen from Chandra/ACIS-I COUP observation, with count
rate from $0.0027$ to $0.0799$ s$^{-1}$, and in off-axis angle
from $0''.35$ to $125''.8$. Both abscissa axes are source number,
sorted with the FWHM of original point sources, before applying
SER but after removing randomization. Furthermore, COUP data
process includes CTI correction (Townsley et al.\ 2002), to reduce
charge transfer problem in ACIS-I CCDs and to recover event grade
information.

The source size, represented by FWHM, was apparently smaller after
applying SER approaches on BI devices, from TSER to SSER, then to
EDSER and to CSDSER, as predicted by BI simulation, demonstrating
the capability to improve the spatial resolution of BI
Chandra/ACIS imaging. At the same time, FI devices illustrate more
modest improvements, by applying SER techniques. The better
performance of static SER than TSER is evident, but from SSER to
EDSER and CSDSER, the improvement is less clear, for the reasons
discussed in section \ref{sec:sim}. However, a small improvements
in effective FI Chandra/ACIS PSF still can be seen after
application of SER techniques.

\clearpage
\clearpage

\begin{figure}
{\centering \resizebox*{4in}{!}{\includegraphics{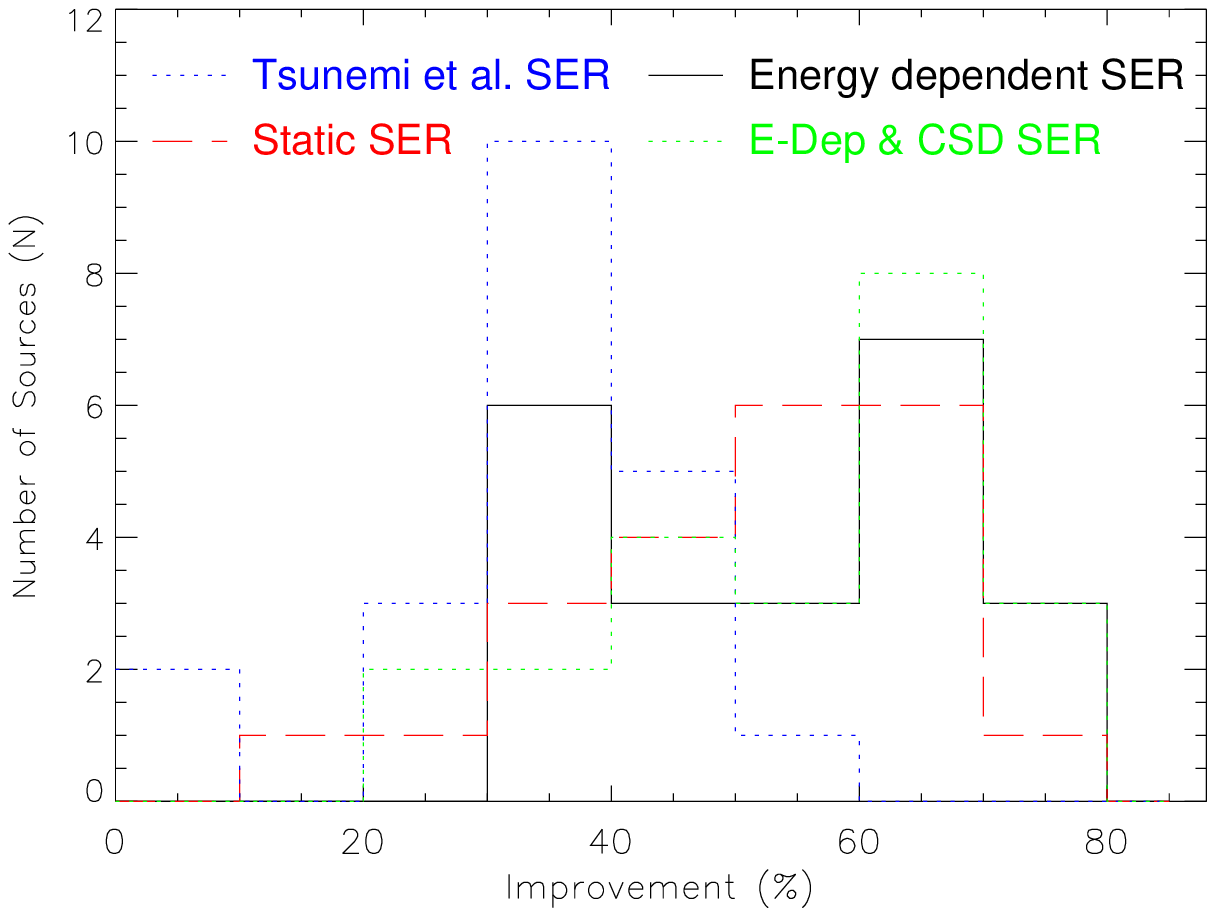}}
\par}
{\centering \resizebox*{4in}{!}{\includegraphics{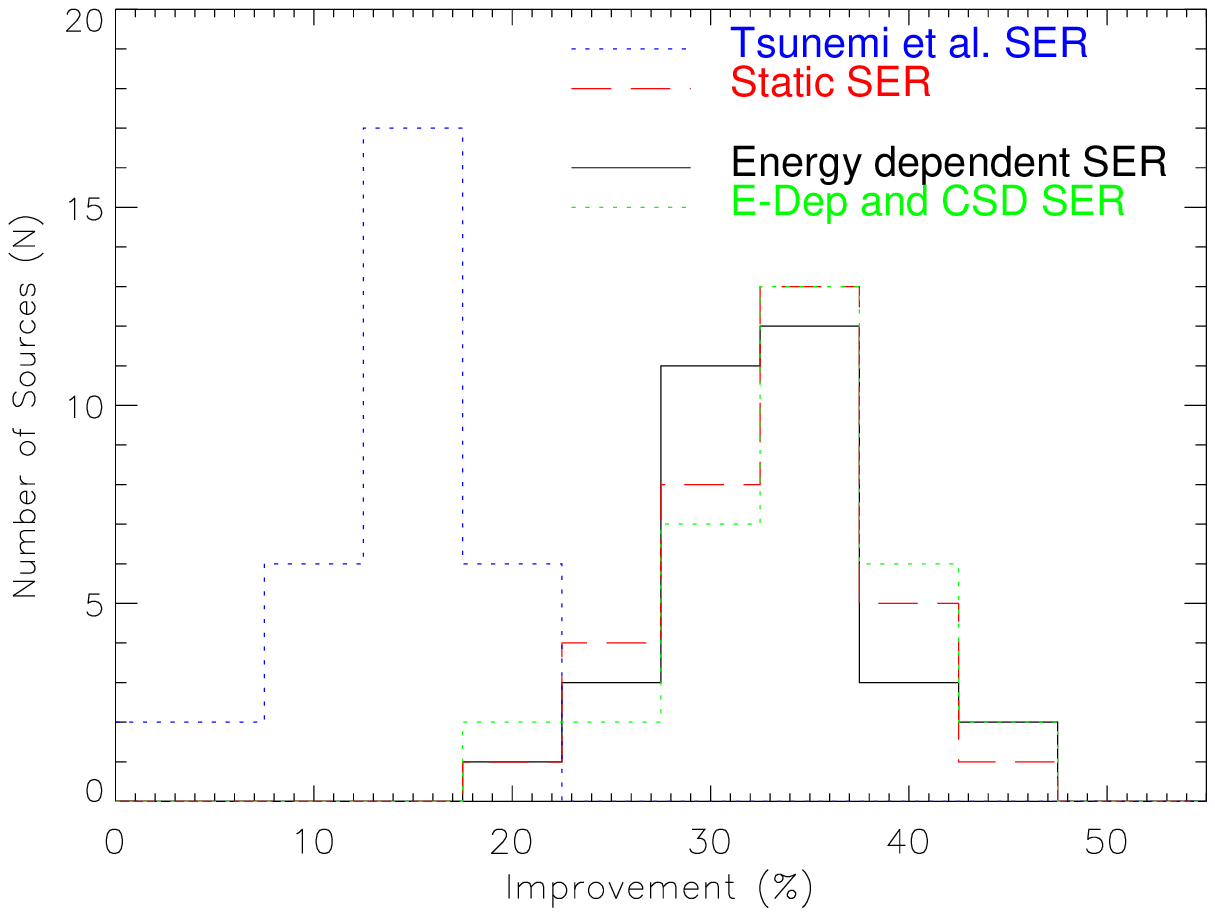}}
\par}
\caption{Comparison of image FWHM improvements using Tsunemi et
al.\ (2001) model, static, energy-dependent SERs, ED and CSD SER
on BI and FI CXO ONC data.} \label{improve}
\end{figure}

Using the definition of improvement given in section
\ref{sec:sim}, we quantitatively evaluate the performance of
different SER methods on ONC data. The top and bottom part of
figure \ref{improve} shows this metric of the improvement for all
SER algorithms for BI and FI Chandra/ACIS sources, respectively.
As expected from MARX simulations, BI data shows superior
improvement for CSDSER and EDSER, while FI data only shows
improvement for modified SERs, and there is no favorite among the
three modified methods. Improvement for most sources in FWHM range
is from 40\% to 70\%, and from 20\% to 50\%, for BI and FI CCDs,
respectively, with the improvement statistically dependent on
off-axis angle.

\clearpage
\clearpage

\section{Summary}
A study of potential improvements to subpixel event repositioning
(SER) for CXO/ACIS data was conducted here, for BI and FI devices.
We formulate modified SER algorithms at three levels of
improvement for both CCD types: (1) inclusion of single-pixel
events and two-pixel split events ({}``static'' SER); (2) in
addition to event grade/split morphology, accounting for the mean
energy dependence of differences between apparent and actual
photon impact positions, based on the results of CCD simulations
({}``energy-dependent'' SER); (3) dependence of the actual PIPs
according to the split charge proportion in the split pixel(s),
event type, and event energy, based on CCD model simulation
results {''charge split dependent'' SER).

All three modified SER methods produce improvements in spatial
resolution over those possible using a static SER algorithm
employing only corner-split events (Tsunemi et al.\ 2001), for
both BI and FI devices. BI and FI CCDs exhibit different
performance and, overall, BI applications benefit more from
angular resolution improvement after applying SER techniques. In
addition, BI data demonstrate the superiority of energy and/or
charge split dependent SER methods, while FI data show only
marginal differences between the various modified SER methods.

\acknowledgements{This research was supported by NASA/CXO grant
G02-3009X to RIT.}

\end{document}